\documentclass[letterpaper,11pt]{article}
\usepackage[british]{babel}
\usepackage[T1]{fontenc}
\usepackage{amssymb,amsmath}
\usepackage[letterpaper,margin=1.1in]{geometry}
\usepackage[margin=0.35in,labelfont=bf,font=small]{caption}
\usepackage{fixmath}
\usepackage[Symbol]{upgreek}
\usepackage[upmu]{gensymb}
\usepackage{tensor}
\ifx\pdftexversion\undefined
  \usepackage[dvips,final]{graphicx}
\else
  \usepackage[pdftex,final]{graphicx}
\fi

\newcommand{\dbar}{{\mathchar'26\mkern-11mu\mathrm{d}}}

\newcommand{\bld}[1]{\boldsymbol{#1}}

\newcommand{\xb}{\boldsymbol{x}}

\newcommand{\cs}{\mathrm{c}}
\newcommand{\ds}{\mathrm{d}}

\newcommand{\ms}{\mathrm{m}}

\newcommand{\rs}{\mathrm{r}}

\newcommand{\Ss}{\mathrm{S}}
\newcommand{\Bcal}{\ensuremath{\mathcal{B}}}

\newcommand{\Lcal}{\ensuremath{\mathcal{L}}}
\newcommand{\Mcal}{\ensuremath{\mathcal{M}}}

\newcommand{\Ocal}{\ensuremath{\mathcal{O}}}

\newcommand{\intfb}[1]{\ensuremath{\int\dbar^4k \,}}
\newcommand{\inttb}[1]{\ensuremath{\int\dbar^3k \,}}

\newcommand{\head}[1]{{\textbf{\footnotesize #1}}}
\newcommand{\ph}[1]{\phantom{#1}}
\newcommand{\ix}[1]{\indices{#1}}

\makeatletter
\def\asterisks#1{\expandafter\@asterisks\csname c@#1\endcsname}
\def\@asterisks#1{%
  \ifcase#1\or *\or **\or ***\or ****\or *****\or ******\or *******\else
  \@ctrerr\fi}

\makeatother

\begin{document}


\pagestyle{headings} \unitlength=1mm

\title{\hfill \hbox{\small{\sffamily{EFI-0518}}} \\\sffamily{\textbf{Cosmological Structure Evolution
and CMB Anisotropies in DGP Braneworlds}}  }
\author{\sffamily{Ignacy Sawicki and Sean M. Carroll} \\
\sffamily{Enrico Fermi Institute, Department of Physics,} \\
\sffamily{and Kavli Institute for Cosmological Physics}\\
\sffamily{University of Chicago, Chicago, IL, USA}}

\maketitle

\abstract{The braneworld model of Dvali, Gabadadze and Porrati (DGP)
provides an intriguing modification of gravity at large distances
and late times.  By embedding a three-brane in an uncompactified
extra dimension with separate Einstein-Hilbert terms for both brane
and bulk, the DGP model allows for an accelerating universe at late
times even in the absence of an explicit vacuum energy.  We examine
the evolution of cosmological perturbations on large scales in this
theory.  At late times, perturbations enter a DGP regime in which
the effective value of Newton's constant increases as the background
density diminishes. This leads to a suppression of the integrated
Sachs-Wolfe effect, bringing DGP gravity into slightly better
agreement with WMAP data than conventional $\Uplambda$CDM. However,
we find that this is not enough to compensate for the significantly
worse fit to supernova data and the distance to the last-scattering
surface in the pure DGP model. $\Uplambda$CDM is, therefore, a
better fit.}

\section{Introduction}
It appears that the acceleration of the expansion of the Universe is
now indubitable: it has been independently corroborated by
measurements of type Ia supernovae \cite{riess:98sn,perlmutter:98sn,
tonry:03sn} and cosmic microwave background radiation observations
by the WMAP satellite \cite{bennett:03wmap}.

The simplest explanation for such an effect is the existence of a
positive cosmological constant. Unfortunately, the estimates for its
natural value are at least 55 orders of magnitude too large (see
\cite{Weinberg:1988cp, carroll:00lambda,peebles:02lambda} for a
review). Another possibility is that the vacuum energy is zero, and
the dark energy comes from the potential of a slowly-rolling scalar
field \cite{Wetterich:1987fm, Ratra:1987rm, Caldwell:1997ii,
Armendariz-Picon:1999rj, Armendariz-Picon:2000dh,
Armendariz-Picon:2000ah, Mersini:2001su, Caldwell:1999ew,
Carroll:2003st, Sahni:1999gb, Parker:1999td}. Yet another
alternative is to modify general relativity so that the effective
Friedmann equation predicts cosmic acceleration even in the absence
of dark energy. \cite{Freese:2002sq, Dvali:2003rk}. A simple
approach along these lines is to add an $R^n$ term to the
Einstein--Hilbert action, with $n<0$, which will affect the
cosmological dynamics in the infrared.\cite{Carroll:2003wy}

The case we will discuss in this work is that proposed by Dvali,
Gabadadze and Porrati (``DGP'') \cite{dvali:00dgp, Dvali:2001gm,
Dvali:2001gx} who suggest that the observed universe might be a
brane embedded in a higher-dimensional space-time, with
Standard-Model fields propagating strictly on the brane. Gravity
propagates in the bulk, but on-brane radiative corrections to the
graviton propagator result in there being induced an additional,
four-dimensional Ricci scalar in the action. The cosmological
solution to this theory with a five-dimensional bulk has been
demonstrated to possess a ``self-accelerated'' branch, as the
universe approaches a de~Sitter phase even when the vacuum energy
vanishes in both the bulk and on the brane. \cite{deffayet:00cosmo}.

By finding the equation governing the growth of radially symmetric
perturbations on a cosmological background with zero cosmological
constants (bulk and brane), Lue \emph{et al.} \cite{lue:04dgpdens}
conclude that the growth of structure is suppressed in a manner
inconsistent with observations and, in the linear regime, the DGP
theory is a theory with a varying gravitational constant.

On the other hand, Ishak \emph{et al}. \cite{Ishak:2005zs} have
simulated data sets for CMB and weak lensing, assuming that the
cosmology is pure DGP, and they have found that, if such data are
analysed assuming that the cosmological acceleration is driven by a
form of dark energy, rather than a modification of gravity, the
forms of best-fit $w$ obtained from the two data sets are
inconsistent.

Attacking the problem with more generality and a using a different
methodology, we show that the quantity which is measured as Newton's
constant for a small perturbation in the Robertson--Walker
background is dependent on the energy-density content of the
universe and is, in general, not equal to the (truly constant)
gravitational constant
driving the Friedmann expansion. This effect results in a change in
the evolution of gravitational potentials in cosmological models
using linear perturbation theory: the matter power spectrum is
somewhat altered and the integrated Sachs-Wolfe (ISW) effect is much
weakened.

It is well-known that the observed CMB anisotropies have less power
on the largest scales than is predicted by the conventional
$\Uplambda$CDM model \cite{bennett:03wmap, deOliveira-Costa:2003pu,
Efstathiou:2003tv}. As a consequence of the suppressed ISW effect,
we find that DGP gravity provides a better fit to the temperature
anisotropies observed by WMAP than does $\Uplambda$CDM. However, DGP
also predicts a somewhat more gradual onset of acceleration than
that expected in $\Uplambda$CDM, which is not as good at matching
the observed Hubble diagram of Type Ia supernovae if we require that
the distance to the last-scattering be fixed. Taken together, we
find that although the low-multipole CMB data is significantly
better fit by DGP, GR is preferred when both CMB and supernova data
are considered for a flat universe.

\section{On-Brane Field Equations}

We start with a single 3-brane embedded in a five-dimensional bulk
$\Mcal$. We will assume that all the standard-model fields are
confined to the brane. Gravity propagates in the bulk and is a fully
five-dimensional interaction, but, as first proposed in the model of
Dvali, Gabadadze and Porrati \cite{dvali:00dgp}, there is an
on-brane radiative correction to the graviton propagator resulting
in the induction of an on-brane, four-dimensional Ricci scalar in
the effective action. Generalizing the DGP model slightly, we will
allow for both a nonzero brane tension, $\lambda$ and bulk
cosmological constant, $\Lambda$.  No fields propagate in the bulk
other than gravity. We thus write down the action: \label{s:defs}
\begin{equation}
    S = \int_\Mcal\!\ds^5 x \sqrt{-g} \left[ \, \frac{^{(5)} R}{2\kappa^2}
    - \Lambda -\frac{K^+}{\kappa^2}-\frac{K^-}{\kappa^2} + \delta(\chi)\left(
    \frac{^{(4)} R}{2\mu^2} - \lambda +\Lcal_\text{SM} \right)\right] \,,
\end{equation}
where $g\ix{_\mu_\nu}$ is the metric in the bulk, $^{(5)}R$ is the
5D Ricci scalar, $^{(4)}R$ is the induced 4D Ricci scalar, $\chi$
parameterizes a vector field such that $\chi=0$ coincides with the
position of the brane everywhere (we assume that such a vector field
exists); a particular choice for this vector field is $n^\mu$,
defined in \eqref{e:ind}. $K^\pm$ is the trace of the extrinsic
curvature of either side of the brane; this term is the
Hawking--Gibbons term necessary to reproduce the appropriate field
equations in a space-time with a boundary. The appropriate energy
scales are represented by $\kappa^2 = 8\uppi
M^{-3}_5$ for the true 5D quantum gravity scale and $\mu^2 = 8\uppi
M^{-2}_4$ for the induced-gravity energy scale on the brane. Finally,
$\Lcal_\text{SM}$ is the standard-model lagrangian, with
fields restricted to the brane.

The ratio of the two gravitational scales, the \emph{cross-over
scale},
\begin{equation}
r_\cs:= \frac{\kappa^2}{2\mu^2} \,,
\end{equation}
was shown by Dvali \emph{et al}. \cite{dvali:00dgp} to appear in the
propagator of the graviton in the 5D-bulk DGP theory. It is at this
distance scale that the potential owing to one-graviton exchange
undergoes a transition from 4D to 5D behavior. Also, Deffayet
\cite{deffayet:00cosmo} has shown that an FRW cosmology enters a
self-accelerated phase when $H \sim r_\cs^{-1}$. If we wish to
explain the cosmological acceleration using this phase, then we need
to set $r_\cs \sim 1$~Gpc, i.e.\ approximately the current Hubble
radius. Dvali \emph{et al.} \cite{dvali:00dgp} show that, at
distances much smaller than the critical radius, when expanded
around the Minkowski background, $\mu^2$ enters the gravitational
potential as the constant of proportionality, i.e.\ $\mu^2 = 8\uppi
G$. This forces us to set $M_4 \equiv M_\text{Pl} = 10^{28}$~eV and,
therefore, $M_5 = 100$~MeV.

Varying the action leads to the following equation of motion:
\begin{equation}
    ^{(5)}G\ix{^\mu_\nu} = \kappa^2 T\ix{^\mu_\nu} \,,
    \label{e:5field}
\end{equation}
where the indices run over all the five dimensions, from 0 to 4. We
write down the induced metric on the brane as
\begin{equation}
    q_{\mu\nu} = g_{\mu\nu} - n_\mu n_\nu \,, \label{e:ind}
\end{equation}
with $n_\mu$ a spacelike vector field with unit norm, normal to the
brane at $\chi = 0$. We can then write down the energy-momentum
tensor as
\begin{equation}
    T_{\mu\nu} = -\Lambda g_{\mu\nu} + S_{\mu\nu}
    \delta(\chi) \, ,
\end{equation}
with the on-brane contribution described by
\begin{equation}
  S_{\mu\nu} = \tau_{\mu\nu} - \lambda q_{\mu\nu}
     - \frac{1}{\mu^{2}} \,^{(4)}G_{\mu\nu} \, .
\end{equation}
Here, $^{(4)}G_{\mu\nu}$ is the Einstein tensor obtained by varying
the usual $\int\!\ds^4 x\,^{(4)}R$, restricted to exist on the
brane, while $\tau_{\mu\nu}$ is the energy-momentum tensor resulting
from varying $\Lcal_\text{SM}$ with respect to $q_{\mu\nu}$. As
already mentioned, the bulk is empty, save for a cosmological
constant $\Lambda$.

We then proceed by following the methodology first proposed by
Shiromizu, Maeda and Sasaki \cite{shiromizu:99proj}: we carry out a
$4+1$ decomposition of the theory \eqref{e:5field} and calculate the
effective on-brane field equations. This result was first obtained
by Maeda \emph{et al}.\ in \cite{maeda:03dgpproj}. The detailed
derivation is presented in Appendix~\ref{s:deriv}.

We use the Gauss equation to calculate the Riemann tensor on the
brane:
\begin{equation}
    ^{(4)}R\ix{^\alpha_{\beta\gamma\delta}} =\
    ^{(5)}R\ix{^\mu_{\nu\rho\sigma}} q^\alpha_{\ph{\alpha}\mu}
    q^\nu_{\ph{\nu}\beta} q^\rho_{\ph{\rho}\gamma}
    q^\sigma_{\ph{\sigma}\delta} + K^\alpha_{\ph{\alpha}\gamma}K_{\beta\delta}
    - K^\alpha_{\ph{\alpha}\delta}K_{\beta\gamma} \label{e:gauss2}
\,,
\end{equation}
where $K_{\mu\nu}:= q^\alpha_{\ph{\alpha}\mu}
q^\beta_{\ph{\beta}\nu} \nabla_\alpha n_\beta$ is the extrinsic
curvature of the brane. We can compute this curvature by assuming a
$Z_2$ symmetry of the codimension; using the Israel junction
conditions for the jumps in the induced metric and the extrinsic
curvature across the brane \cite{israel:66junc}, we obtain
\begin{align}
    2q_{\mu\nu}^+ &= q_{\mu\nu}^+-q_{\mu\nu}^- =: [q_{\mu\nu}] = 0 \\
    2K_{\mu\nu}^+ &= [K_{\mu\nu}] = -\kappa^2
    \left(S_{\mu\nu}-\frac{1}{3}q_{\mu\nu}S \right) \,.
\end{align}
We now contract \eqref{e:gauss2} appropriately and, after some
manipulation, we obtain the equation for the 4D Einstein tensor:
\begin{align}
   \left(1+\frac{\lambda\kappa^4}{6\mu^2}\right)\ ^{(4)} G_{\mu\nu} &=
   -\left(\frac{\kappa^2}{2}\Lambda +\frac{\kappa^4\lambda^2}{12}\right)
   q_{\mu\nu}+ \frac{\lambda
   \kappa^4}{6}\tau_{\mu\nu} + \frac{\kappa^4}{\mu^4} f_{\mu\nu} - E_{\mu\nu}
   \,, \label{e:me2}
\end{align}
where the tensors $f_{\mu\nu}$ and $E_{\mu\nu}$ are defined as
\begin{align}
\label{fmunu} f_{\mu\nu} &:= \frac{1}{12}A A_{\mu\nu} -
    \frac{1}{4}A\ix{_\mu^\alpha}A_{\nu\alpha} +
    \frac{1}{8}q_{\mu\nu} \left(A_{\alpha\beta}A^{\alpha\beta}-
    \frac{1}{3}A^2 \right) \\
E_{\mu\nu} &=\
^{(5)}C\ix{^\perp_\alpha_\perp_\beta}q\ix{^\alpha_\mu}q\ix{^\beta_\nu}
\, ,
\end{align}
with
\begin{equation}
  A_{\mu\nu} :=\ ^{(4)} G_{\mu\nu}-\mu^2\tau_{\mu\nu}\, .
\end{equation}
$E_{\mu\nu}$ is the projection onto the brane of the bulk Weyl
tensor $^{(5)}C\ix{^\alpha_{\beta\gamma\delta}}$ ($\perp$ signifies
an index contracted with $n_\mu$).
$f_{\mu\nu}$ describes brane terms that are quadratic in $\tau_{\mu\nu}$
and $G_{\mu\nu}$; henceforth we will drop the superscript $^{(4)}$ and
assume all quantities are four-dimensional unless otherwise specified.
In Appendix~\ref{s:deriv} we explicitly define the cross terms
appearing in $f_{\mu\nu}$ as $\pi_{\mu\nu}$,
$\gamma_{\mu\nu}$ and $\xi_{\mu\nu}$; see equations
\eqref{e:pi}---\eqref{e:xi} for precise definitions.

We can recover the standard Einstein equations from \eqref{e:me2} by
sending the brane tension $\lambda$ to infinity, i.e.\ in the limit
when the brane is `stiff'.

The difference between DGP gravity and the standard Randall--Sundrum
result of \cite{shiromizu:99proj} is the presence of the 4D Einstein
tensor in $S_{\mu\nu}$, resulting in additional terms containing
$^{(4)}G_{\mu\nu}$ explicitly: we now have an equation which is
quadratic in the Einstein tensor, i.e.\ curvature can act as a
source of curvature, leading to a potentially non-zero Ricci in the
vacuum (as obtained, for example, by Gabadadze and Iglesias in
\cite{gabadadze:04schwarz,Gabadadze:2005qy} for their
Schwarzschild-like solution in DGP).

\section{Cosmological Solution}
\label{cosmology}

The quadratic nature of \eqref{e:me2} makes this formulation
computationally imposing. In addition, it was already noted by
Shiromizu \emph{et al}.\ in \cite{shiromizu:99proj} that the
transverse-traceless component of $E_{\mu\nu}$ contains the
information about gravitational radiation coming off and onto the
brane as well as the transition from 4D to 5D gravity, and its
evolution is necessarily dependent on the state of the bulk: the
equation of motion for $E_{\mu\nu}$ does not close on the brane.

Nevertheless, the large symmetry of a Robertson-Walker universe
allows us to make progress. Choosing Gaussian normal co-ordinates,
with the brane positioned at $y=0$ and a Robertson--Walker metric on
the brane, our bulk metric becomes:
\begin{equation}
    ds^2 = -N^2(y) dt^2 + A^2(t,y) \gamma_{ij} dx^i dx^j + dy^2
    \,,
\end{equation}
and we are allowed to pick the normalization such that $N(0)=1$ by
rescaling the time variable; $\gamma_{ij}$ is a 3D
maximally-symmetric spatial metric. We also define the value of $A$
on the brane: $a(t):= A(t,0)$. In these coordinates, the 4D tensors
are significantly simplified, with only zero entries in the columns
and rows corresponding to the dimension perpendicular to the brane.
We thus obtain for $G\ix{^\mu_\nu}$:
\begin{align}
    G\ix{^0_0} &= -3\left( \frac{\dot{a}^2}{a^2} + \frac{k}{a^2}
    \right) \\
    G\ix{^i_j} &= -\left(2\frac{\ddot{a}}{a}+\frac{\dot{a}^2}{a^2} +
    \frac{k}{a^2}\right) \delta\ix{^i_j}\\
    G\ix{^\mu_\mu} &= -6\left(
    \frac{\ddot{a}}{a}+\frac{\dot{a}^2}{a^2}+\frac{k}{a^2} \right)
\,,
\end{align}
with the dot representing differentiation with respect to the $t$
coordinate. In addition, we assume that the brane is filled with a
homogeneous distribution of a perfect fluid, such that
\begin{align}
    \tau\ix{^\mu_\nu} &= \text{diag}(-\rho,p,p,p,0)\ \\
    p &= w\rho \,.
\end{align}
%
%
This allows us to obtain the 0-0 component of the quadratic tensor
$f^\mu{}_\nu$ of (\ref{fmunu}):
\begin{equation}
  f^0{}_0 = -\frac{1}{12}\left[\mu^2\rho - 3\left(\frac{{\dot a}^2}{a^2}
  + \frac{k}{a^2}\right)\right]^2 \, .
\end{equation}

The remaining issue is the tensor $E_{\mu\nu}$. Since we are dealing
with an isotropic and homogeneous universe, we must set it to be
just a function of the time on the brane. Since it is traceless, it
behaves just like radiation and decays as $a^{-4}$; because of this
property, this term is usually referred to as dark radiation (see,
for example, the review of Maartens \cite{Maartens:2003tw}). For the
moment we will set:
\begin{equation}
    E\ix{^0_0} = \frac{C}{a^4} \, ,
\end{equation}
although ultimately we will set this term to zero.

Substituting the above into \eqref{e:me2} and replacing
$\frac{\dot{a}}{a}$ with $H$, we obtain a quadratic equation for the
energy density (or, equivalently, for $H^2$):
\begin{align}
    \frac{\kappa^4}{12}\rho^2 + \left( \frac{\lambda\kappa^4}{6} -
    \frac{\kappa^4}{2\mu^2} \left( H^2 + \frac{k}{a^2} \right)
    \right) \rho +\frac{3}{4} \frac{\kappa^4}{\mu^4} \left(H^2 +
    \frac{k}{a^2}\right)^2 - \notag \\
 \qquad - 3\left( 1+
    \frac{\kappa^4\lambda}{6\mu^2}\right)
    \left( H^2 + \frac{k}{a^2} \right) + \frac{\kappa^2}{2}
    \left(\Lambda + \frac{\kappa^2 \lambda^2}{6} \right) + \frac{C}{a^4} =
    0 \label{e:mess} \,,
\end{align}
where $k=-1,1,0$ depending on whether the spatial hypersurface on
the brane has negative, positive or no curvature. We can write this
relation as a version of the Friedmann equation with modified
dependence on the Hubble parameter,
\begin{equation}
    H^2 \pm \frac{2\mu^2}{\kappa^2} \sqrt{ \left(
    H^2+ \frac{k}{a^2} \right) - \frac{\kappa^2}{6}\Lambda -
    \frac{C}{3a^4}} = \frac{\mu^2}{3}(\rho + \lambda) -
    \frac{k}{a^2} \,.
\end{equation}
So, provided that the $\Lambda$ and $C$ terms remain negligible, and
$H^2 + \frac{k}{a^2} \gg 2\frac{\mu^2}{\kappa^2}$, the evolution of
the scale factor in the early universe does not differ from that in
standard FRW cosmology. As $H$ decreases the evolution becomes
non-standard; this can be demonstrated more clearly by solving
\eqref{e:mess} for $H^2$:
\begin{equation}
    H^2 = 2\frac{\mu^4}{\kappa^4} + \frac{\mu^2}{3}
    (\rho+\lambda) - \frac{k}{a^2} + 2\epsilon \frac{\mu^2}{\kappa^2}
    \sqrt{\frac{\mu^4}{\kappa^4}+\frac{\mu^2}{3}(\rho+\lambda) -
    \frac{\kappa^2}{6}\Lambda - \frac{C}{3a^4}} \label{e:friedmann}
\,,
\end{equation}
with $\epsilon=\pm 1$ representing the two possible embeddings of
the brane in the bulk (see \cite{deffayet:00cosmo}). The above
result has been obtained previously obtained by Collins and Holdom
\cite{collins:00brane} and Shtanov \cite{shtanov:01fried}.

We are going to follow Deffayet \cite{deffayet:00cosmo} by naming
the $\epsilon=+1$ branch as `self-accelerated' and $\epsilon=-1$
branch as `non-accelerated'. The reason for this nomenclature
becomes obvious in the case of flat bulk and a zero brane tension
and no dark radiation, i.e.\ $k,\lambda,\Lambda,C = 0$. Equation
\eqref{e:friedmann} then reduces to that originally obtained by
Deffayet (\emph{ibid}):
\begin{equation}
    H^2 = 2\frac{\mu^4}{\kappa^4} + \frac{\mu^2}{3}\rho + 2\epsilon
    \frac{\mu^2}{\kappa^2}\sqrt{\frac{\mu^4}{\kappa^4}+\frac{\mu^2}{3}\rho}
\,.
\end{equation}
As mentioned previously, this solution contains a cross-over scale
above which the self-acceleration term dominates:
\begin{equation}
    r_\cs := \frac{\kappa^2}{2\mu^2} \,. \label{e:rcdef}
\end{equation}
For $\epsilon=+1$, once $\mu^2\rho/3 \ll r_\cs^{-2}/2$, $H$
approaches the nonzero constant $r_\cs^{-1} = 2\mu^2/\kappa^2$, and
the universe enters an accelerated de~Sitter phase. If we wish to
use this model to replace the effects of the cosmological constant,
we should set $r_\cs$ to be approximately the current Hubble radius.
The non-accelerated branch with $\epsilon=-1$ behaves like the usual
Friedman universe, with $H^2$ tending to zero in the flat-universe
case.

Note that there are two distinct limits in which we can recover the
ordinary Friedmann equation in the presence of a cosmological constant,
\begin{equation}
  H^2 = \frac{\mu^2}{3}(\rho + \rho_\text{vac}) - \frac{k}{a^2}\, ,
\end{equation}
where $\rho$ is the energy density in everything other than the
cosmological constant.  One limit is to simply let $\kappa^2
\rightarrow \infty$, decoupling the extra dimension entirely and
leaving us with $\rho_\text{vac} = \lambda$.  The other is to
set the brane tension to zero, $\lambda = 0$, and take both $\kappa^2$
and $-\Lambda$ to infinity while keeping their ratio constant,
yielding
\begin{equation}
  \rho_\text{vac} = \sqrt{-\frac{6\Lambda}{\kappa^2}}\, .
\end{equation}
In our investigation of DGP cosmology, we will set the brane tension
to zero while leaving the bulk cosmological constant $\Lambda$ as a
free parameter and calculating its likelihood as determined by the
data.  Then $\Lambda =0$ corresponds to ``pure DGP,'' while $\Lambda
\rightarrow -\infty$ corresponds to ordinary $\Uplambda$CDM.

\section{Linearized Equations and the Potential}
\label{s:lin}

In this section, we will derive the Poisson equation
for a perturbation of the Robertson--Walker background and
demonstrate that it deviates from the usual version: in its linear
regime, the DGP theory on the brane is a varying-$G$
theory. This derivation will assume that we are able to
neglect all terms not linear in the gravitational potential. This
range of validity of this assumption is discussed in \S\ref{s:ass}.

We start off by introducing scalar perturbations to the metric.
Since we are only going to be dealing with on-brane directions, we
can use the 4D formalism in the conformal Newtonian gauge, parameterizing the
perturbed metric by:
\begin{equation}
    ds^2 = -(1+2\Psi(\xb,t))dt^2 + a(t)(1+2\Phi(\xb,t))\gamma_{ij} dx^i
    dx^j \,.
\end{equation}
This allows us to calculate the Einstein tensor up to first order in
the perturbations:
\begin{equation}
    G\ix{^0_0} = -3H^2 + 6H^2\Psi -6H\dot{\Phi} + \frac{2}{a^2}
    \bld{\nabla}^2\Phi + \Ocal(\Phi^2, \Psi^2) \,.
\end{equation}
The above approximation is permitted provided $\Phi,\Psi \ll 1$. If
we consider distance scales much smaller
than Hubble scale, i.e.\ upon taking the Fourier transform,
$(k/a)^2\Phi \gg H^2\Psi, H\dot{\Phi}$, then from the first-order
terms we recover the potential for the flat Minkowski spacetime:
\begin{equation}
    ^{(1)}G\ix{^0_0} = \frac{2}{a^2} \bld{\nabla}^2\Phi
    \label{e:minke} \,.
\end{equation}
Alternatively, by including the Hubble flow, we can obtain the 0-0
component of the cosmological evolution equation for the
gravitational potentials in GR:
\begin{equation}
    \frac{1}{a^2} \bld{\nabla}^2 \Phi +3H^2\Psi - 3H\dot\Phi =
-\frac{\mu^2}{2}\bar\rho \delta \label{e:grcos}
\end{equation}
where $\bar{\rho}$ is the average matter/radiation energy density of
the universe, $\delta\rho$ is the deviation from this mean, and the
fractional density excess is defined as
$\delta:=\delta\rho/\bar{\rho}$.

In DGP gravity, the 0-0 component of the right-hand side of the modified
Einstein equation \eqref{e:me2}, first-order in $\Phi$ or $\Psi$, is:
\begin{align}
    ^{(1)}[\text{RHS}]\ix{^0_0} &= -\delta\rho \frac{\kappa^4}{6}
    \left(\lambda+ \bar\rho \right)
    + \frac{\kappa^4}{\mu^2} H\dot\Phi - \frac{3\kappa^4}{\mu^4}
    \left(H^3\dot\Phi +H^4\Psi \right) + \frac{\kappa^4}{2\mu^2}
    (\delta\rho - 2\bar\rho\Psi)H^2 \notag \\
    &\qquad+ \left(\frac{\kappa^4}{\mu^4}H^2 -
    \frac{\kappa^4\bar\rho}{3\mu^2}\right)a^{-2}\bld\nabla^2\Phi \,.
\end{align}

In calculating the above, we have assumed that the effect of the
perturbations of the Weyl tensor, $E_{\mu\nu}$ in this equation at
sub-horizon scales are insignificant. Since we are dealing with a
quasi-static situation, gravitational radiation is negligible. In
addition, this tensor encodes the transition of gravity from 4D to
5D behavior. However, this effect only occurs at a length scale
determined by $r_\cs$, which, as shown in the data fits of
\S\ref{s:fits}, is much larger than the horizon scale even today,
let alone in the past. As such, it is unlikely to have any
significant effect on the quantities under consideration.

Performing some algebraic manipulation and substituting for $H^2$
from \eqref{e:friedmann} we obtain the DGP equivalent of
\eqref{e:grcos}:

\begin{equation}
    \frac{1}{a^2} \bld{\nabla}^2 \Phi +3H^2\Psi - 3H\dot\Phi =
-\frac{\mu^2}{2}\bar\rho \left( 1+ \frac{\epsilon}{\sqrt{1+
\frac{4}{3} \mu^2 r_\cs^2 \left(\bar{\rho} + \lambda -r_\cs \Lambda
-\frac{\kappa^4 C}{\mu^4 a^4}\right)}}\right)\delta \label{e:dgpcos}
\end{equation}
If we apply similar approximations to those that led to
\eqref{e:minke}, i.e.\ assuming that $\Phi, \Psi \ll 1$, drop terms
containing $H$ by considering scales over which the Hubble flow is
negligible, we obtain:
\begin{equation}
    \bld{\nabla}^2\Phi = -\frac{\mu^2\bar\rho a^2}{2} \left( 1+ \frac{\epsilon}{\sqrt{1+
\frac{4}{3} \mu^2 r_\cs^2 \left(\bar{\rho} + \lambda -r_\cs \Lambda
-\frac{\kappa^4 C}{\mu^4 a^4}\right)}}\right) \delta \label{e:poi}
\,.
\end{equation}
This form of \eqref{e:poi} signifies that the weak-field
approximation in a Robertson--Walker background is a varying-$G$
theory, with the effective Newton's constant dependent on the
average energy density of the Universe:
\begin{equation}
    G_\text{eff} = \frac{\mu^2}{8\uppi} \left(1+
    \frac{\epsilon}{\sqrt{1+\frac{4}{3}\mu^2 r_\cs^2 \left(\bar\rho+\lambda-
    r_\cs\Lambda- \frac{3 C}{a^4 \mu^2}\right)}} \right)
    \label{e:geff} \,.
\end{equation}
Note that, in the numerical solutions described below, we do {\em not}
neglect terms containing $H$; we are taking that limit here purely for
expository purposes.

The large-scale evolution of the universe is always driven by
\eqref{e:friedmann}, i.e.\ the energy scale encoded in $\mu^2$. This
is the parameter which drives the conditions during, for example,
nucleosynthesis. However, at least in part, the evolution of
structure is driven by the effective Newton's cons\-tant, i.e.\
equation \eqref{e:geff}: this is generally true at late times for
diffuse clouds of gas. This insight will provide a constraint for
some of the parameters of the theory.

As we can observe from equation~\eqref{e:geff} and assuming that we
are in the self-accelerated branch of the solution ($\epsilon =
+1$), the value of the effective $G_\text{eff}$ varies from one to,
at most, two times the underlying Newton's constant, with the
increase occurring in the late universe as $\bar\rho \rightarrow 0$.
(A related effect has been observed in the presence of
Lorentz-violating vector fields \cite{Carroll:2004ai}, which result
in a cosmology where Newton's constant differs from the constant
relating energy density to the Hubble parameter in the Friedmann
equation.)  In the limit described at the end of
section~\ref{cosmology}, in which we take $\Lambda \rightarrow
-\infty$, the time-dependent piece of $G_\text{eff}$ goes away, and
we obtain $G_\text{eff}=\mu^2/8\uppi =$~constant, so that ordinary
$\Uplambda$CDM is indeed recovered.

By considering the $i$--$j$ equations of motion in GR, we can get
the an equation relating the two potentials $\Phi$ and $\Psi$ to the
anisotropic stress of the cosmological fluid, $\pi$:

\begin{equation}
    \frac{k^2}{a^2} (\Phi + \Psi) = \mu^2 \pi
\end{equation}
Since the anisotropic stress is negligible in models with no
neutrinos, we can set $\Phi = -\Psi$. The analogous calculation in
DGP is a little more complex; however, it yields similar results:
without a source of significant anisotropic stress in the
cosmological fluid, we can set $\Phi = -\Psi$.

\begin{align}
\left( 1 + \frac{\kappa^2}{6\mu^2}\left( \lambda +
\frac{1}{2}\bar\rho(1-3w) - \frac{3\ddot{a}}{\mu^2 a}\right)\right)
\frac{k^2}{a^2} (\Phi+\Psi) =\\= \frac{\kappa^2}{6\mu^2} \left(
\lambda + \frac{1}{2}\bar\rho(1-3w) - \frac{3\ddot{a}}{\mu^2 a}
\right) \mu^2 \pi - \pi_E \notag
\end{align}

$\pi_E$ is a new term and is the anisotropic stress encoded in the
the Weyl tensor $E_{\mu\nu}$, i.e.\ a result of off-brane effects,
such as graviton evaporation into the bulk and any gravitational
waves going off or coming onto the brane. We will set this to zero
in the subsequent analysis.

\section{Range of Validity of Linear Regime}
\label{s:ass}

In section \ref{s:lin}, we assumed that we can use the
linear approximation to the modified Einstein's equation
\eqref{e:me2}. It is important to verify this explicitly, since the
large magnitude of the coefficients of the quadratic terms in
\eqref{e:me2}, might lead to their dominating over the linear terms.

After a systematic review of all the quadratic terms in the
expansion of \eqref{e:me2}, we conclude that the quadratic term
which is most likely to be large arises from the
$(G\ix{_\mu_\nu})^2$ term and is of the form
\begin{equation}
    \frac{\kappa^4}{\mu^4} \left(\bld{\nabla}^2 \Phi\right)^2
\end{equation}
Assuming that the Poisson equation is approximately valid, despite
the evolving cosmo\-logical background, we have $\bld\nabla^2\Phi
\sim \mu^2 \bar\rho \delta$, and using the definition of $r_\cs$,
\eqref{e:rcdef}, leads to the condition that for the linearization
of DGP to be valid
\begin{equation}
    r^2_\cs \mu^2 \bar\rho \delta \ll 1
\end{equation}
If we assume that the density perturbation is in the form of a
spherical top hat with a characteristic size $D$, we can write down
the mass of this object as $M \sim D^3 \bar\rho \delta$. Finally,
the Schwarzschild radius in four dimensions is $r_\Ss \sim \mu^2 M$,
giving us:
\begin{equation}
    r^2_\cs r_\Ss \ll D^3
\end{equation}

We have thus recovered the result first discovered by Dvali et al
\cite{dvali:2002vf}, that there is a new scale in DGP theory,
\begin{equation}
  r_* = (r_\cs^2 r_\Ss)^{1/3}\, .
\end{equation}
At small distances (between the Schwarzschild radius and $r_*$),
gravity behaves essentially as in 4D GR, because
the quadratic terms dominate the modified Einstein
equation \eqref{e:me2}. If we set $\Lambda = \lambda = 0$, in this
regime we are looking for the solution to
\begin{equation}
    f\ix{^\mu_\nu} \approx 0
\end{equation}
with the obvious solution being $G\ix{^\mu_\nu} = \mu^2
\tau\ix{^\mu_\nu}$, i.e. the usual 4D Einstein equation.
On intermediate scales (between $r_*$ and $r_\cs$), gravity is
described by a scalar-tensor theory, consistent with our description
above in terms of a time-dependent gravitational constant.  It is in
this regime that the linearized DGP equations are valid.

As we will see below, as a perturbation of fixed comoving size
expands along with the universe, it typically goes from being less
than $r_*$ to being greater than $r_*$.  Each mode
is therefore described by 4D GR at early times, and later on by
linearized DGP.

\section{Cosmological Simulations}
\subsection{The Simulation}

In order to test the effects on cosmological observations of
modifying gravity, we built a simple cosmological simulation
containing only dark matter and radiation and modeled their
evolution according to linearized equations -- linearized DGP
or ordinary 4D GR, depending on the regime a given mode
is in. The model produces as
outputs the matter power spectrum and the contribution of the
Sachs--Wolfe effect (both integrated and non-integrated) to the
radiation power spectrum. The difference between the matter power
spectra is slight; however, the impact on the ISW is significant,
resulting in a large reduction of power at low multipoles.

We compare the results of the concordance $\Uplambda$CDM model to
that of DGP cosmology.  The concordance model is defined by
$\Omega_\ms = 0.27$, $\Omega_\ms/\Omega_\rs = 3234$ and
$\Omega_\lambda + \Omega_\ms + \Omega_\rs = 1$. For the DGP
cosmologies we use the same matter to radiation energy ratio, and
choose the self-accelerated embedding, $\epsilon = +1$.  We keep
$\Omega_\rs h^2$ constant. As our aim is to explain the acceleration
of the expansion of the universe using effects arising from modified
gravity, we have set the brane tension $\lambda$ (which is
equivalent to the cosmological constant in $\Uplambda$CDM) to zero
in all our considerations.  For simplicity we have also set $C$ and
$k$ to zero, i.e.\ we are assuming a flat cosmology. We keep the
bulk cosmological constant $\Lambda$ as a free parameter.

We parameterize the DGP models by rewriting \eqref{e:friedmann} as:
\begin{equation}
    \frac{H^2}{H_0^2} = \frac{1}{2\beta^2} + \Omega_\ms a^{-3}+\Omega_\rs a^{-4} +\frac{1}{\beta}
\sqrt{\frac{1}{4\beta^2}+\Omega_\ms a^{-3} + \Omega_\rs a^{-4} -
\Omega_\Lambda} \label{e:modfried}
\end{equation}
where $H_0$ is the Hubble parameter today,
\begin{align*}
    &\beta := r_\cs H_0 \qquad &\text{cross-over radius in Hubble
units} \\
    &\Omega_\ms +\Omega_\rs := \mu^2\bar\rho_0/3H_0^2 = 1-\beta^{-1}\sqrt{1-\Omega_\Lambda}
 & \text{energy density}\\
&\Omega_\Lambda := \kappa^2\Lambda/6H_0^2 \qquad &\text{dimensionless bulk cosmological constant} \\
&\Omega_\rs = \Omega_\ms/3234 \qquad &\text{fractional radiation
density}
\end{align*}
$\Omega_\ms$ is the contribution of matter to the total energy
density of the universe (the remainder coming from the DGP
curvature). Realistic DGP models will have $\Omega_\ms \sim 0.3$,
just as in $\Uplambda$CDM. The energy scale for the bulk
cosmological constant is $\Lambda \approx (10^{-8}~\text{eV})^5
\Omega_\Lambda$.

\begin{figure}[htp]\begin{centering}
\includegraphics[width=10cm]{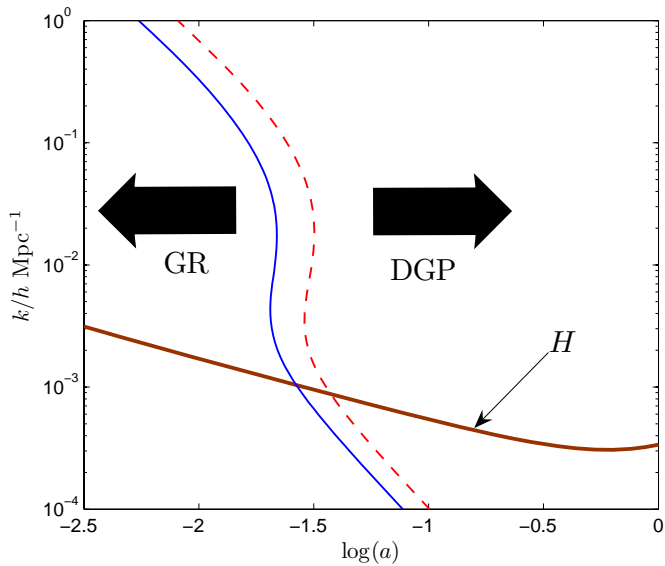}
\caption{Value of scale factor at which the gravity driving
evolution transitions from GR to linear DGP. Solid line is for
Minkowski bulk ($\Omega_\Lambda = 0$), dashed for $\Omega_\Lambda =
-1$. Shape of transition line is a processed dark-matter power
spectrum. \label{f:dgptrans}}
\end{centering}\end{figure}

A significant issue in modeling DGP is deciding on where the the
transition between the Einstein and the linear DGP regimes occurs
and how to implement it. The simulation was built to switch from GR
to linearized DGP instantaneously at the point in evolution defined
by the scale $r_*$. We will apply DGP gravity when the scale of the
perturbation is
\begin{equation}
    D^3 \sim r_*^3 =  2GM r_\cs^2 = \frac{1}{24} \mu^2 D^3 r_\cs^2
\bar\rho\delta
\end{equation}
The perturbation size is eliminated from the relationship, yielding
for the fractional density perturbation at which GR transitions to
linear DGP:
\begin{equation}
    \delta \sim \frac{24}{\mu^2 r_\cs^2 \bar\rho_0} a^{3(1+w(a))}
\end{equation}
where $w(a)$ is the effective equation of state parameter for the
fluid comprising the universe and $a$ is the scale factor; $\delta$
is dependent on the initial conditions, and thus is a function of
the mode under consideration. Thus, for matter and radiation, the
transition occurs from GR to DGP and occurs only once in the
evolution of the perturbations. This transition has been plotted on
Figure~\ref{f:dgptrans}. We can determine the success of the
splicing between GR and the linear DGP regimes by comparing the
value of the effective Newton's constant at the transition point. We
found that the transition occurs at around $z = 20$ for modes of the
size of the horizon today, with $G_\text{eff} \approx 1.02 G$, with
this transition occurring at larger redshifts for higher modes and
$G_\text{eff}$ even closer to 1.

\begin{figure}[htp]\begin{centering}
\includegraphics[width=10cm]{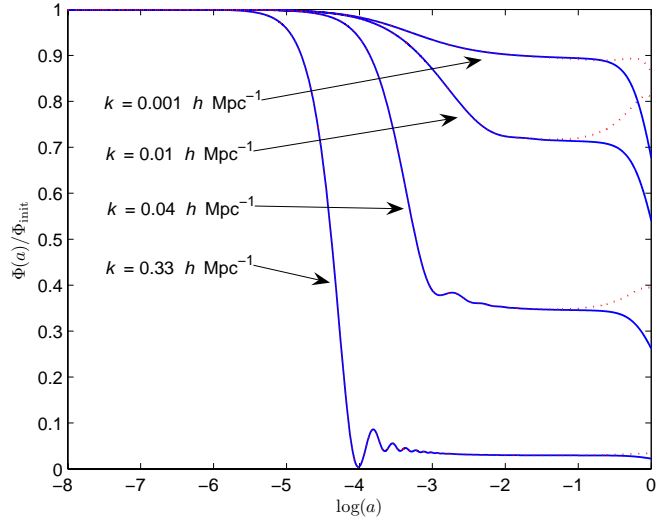}
\caption{Evolution of Newtonian potential $\Phi$ in GR and DGP
(GR--- solid lines, DGP---dashed). In DGP, the growth of the
effective Newton's constant leads to a wavelength-dependent growth
in the potentials at late times, as opposed to the decay observed in
$\Uplambda$CDM.\label{f:pots} ($\Uplambda$CDM: concordance model;
DGP: $\Omega_\Lambda = 0, \beta = 1.38$)}
\end{centering}\end{figure}

\begin{figure}[htp]\begin{centering}
\includegraphics[width=10cm]{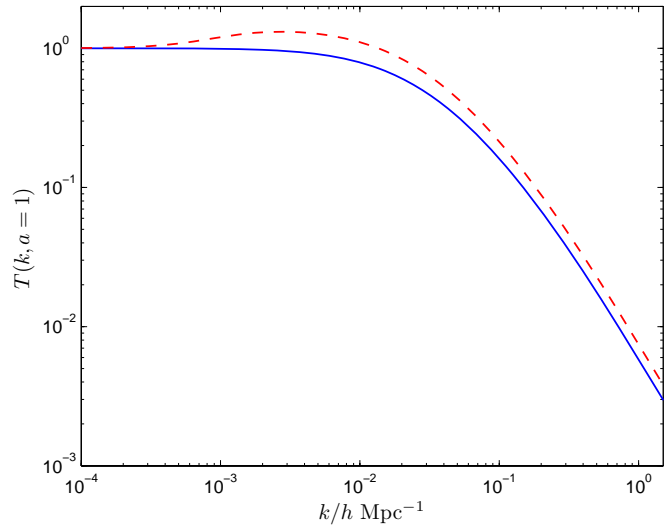}
\caption{Transfer functions in $\Uplambda$CDM and DGP
($\Omega_\Lambda = 0$, $\beta = 1.38$). The transfer function is
defined here as ratio of size of initial perturbation in $\Phi$ to
its final value, renormalized to 1 at large scales. In GR, the
growth rate for the potential during dark-energy domination is
independent of $k$, so the quantity shown here is equivalent to the
usual transfer function. For DGP, the growth rate depends on $k$.
\label{f:transfer}}
\end{centering}\end{figure}

In Figure~\ref{f:pots} we plot the evolution of the Newtonian
gravitational potential $\Phi$ for different modes in both
$\Uplambda$CDM and DGP.  Each mode diminishes in amplitude when it
first comes into the Hubble radius (more so during radiation
domination), before settling to a constant amplitude.  In
$\Uplambda$CDM, linear modes once again decrease in amplitude when
vacuum energy begins to dominate over matter.  In DGP, in contrast,
the effective gravitational constant increases just when the
universe begins to accelerate, leading to an increase in the
amplitude of $\Phi$ that can be appreciable for certain wavelengths.
This behavior is also reflected in the transfer function for dark
matter perturbations, plotted in Figure~\ref{f:transfer}, which
shows a slight enhancement in DGP over $\Uplambda$CDM.

\subsection{Constraints from Expansion History}
\label{s:fits}

Before considering details of CMB anisotropies, we turn to two sources of
constraints on the expansion history of the universe:  the Hubble
diagram of Type~Ia supernovae, and the distance to last scattering
as measured by WMAP.  These imply a tight relationship between the
two free parameters in the DGP model:  $\beta \equiv r_c H_0$ and
$\Omega_\Lambda$.  We will then calculate CMB anisotropies in models
that obey this relationship.

For the supernova data, we used the Riess \emph{et al}. Gold SNe Ia
data set \cite{Riess:2004nr} (156 supernovae) and searched for the
parameters minimizing $\chi^2$. In the DGP scenario with both
$\beta$ and $\Omega_\Lambda$ free, the optimization routine chooses
very large and negative values for $\Omega_\Lambda$, corresponding
to the $\Uplambda$CDM limit as discussed at the end of
Section~\ref{cosmology}. The supernova data, in other words, prefer
a very large and negative bulk cosmological constant, reducing the
observable physics to GR. This is the result of the fact that
$\Uplambda$CDM fits the supernova data slightly better (see
Table~\ref{t:bestfits}). In terms of the effective equation-of-state
parameter $w$, the supernova data prefer $w\approx -1$ or even a
little less, while pure DGP corresponds to $w_\text{eff}\approx
-0.7$ today.

If we fix $\Omega_\Lambda = 0$, supernova data fit best when $\beta
= 1.20$, implying $\Omega_\ms = 0.20$. The details of the fit are
presented in Table~\ref{t:bestfits}.

The WMAP experiment has determined, to a high level of precision,
the distance to the last-scattering surface as $d_\text{ls} =
14.0^{+0.2}_{-0.3}$~Gpc~$ \equiv 3.32^{+0.04}_{-0.08} \, H_0^{-1}$
\cite{bennett:03wmap}. This restriction needs to be included in the
likelihood calculation. Whereas for $\Uplambda$CDM the two sets of
data are consistent, they are \emph{not} so for pure DGP. A good fit
for the CMB distance in DGP with $\Omega_\Lambda = 0$ requires a
higher $\beta = 1.66$, implying $\Omega_\ms= 0.40$.

This is very different from the requirements of the supernovae.
Putting the two sets of data together results in the conclusion that
owing to the \emph{inconsistency} between parameters preferred by
the SN and CMB distance data, the overall fit to pure DGP assuming a
flat cosmology ($\Omega_\Lambda = 0$) is much worse than for
concordance $\Uplambda$CDM.

\begin{table}[htp]\begin{centering}
\begin{tabular}{lrrrr}
    \hline\hline
    \head{Scenario} & \head{$\beta\equiv r_\cs H_0$}& \head{$\Omega_\ms$}&
         \head{$\chi^2$ per d.o.f.} &\head{Confidence}\\
    \hline
    SN DGP & $1.26^{+0.01}_{-0.02}$ & $0.20 \pm 0.01$ &
        1.15 & 9\% \\
    SN $\Uplambda$CDM & --- & 0.30 & 1.14 & 10\% \\
    CMB dist. DGP & $1.66^{+0.08}_{-0.02}$ & $0.40^{+0.02}_{-0.01}$ &
        --- & --- \\
    CMB dist. $\Uplambda$CDM & --- & 0.29 & --- & --- \\
    Total DGP & $1.38^{+0.02}_{-0.01}$ & $0.28\pm0.01$ & 1.21 & 4\% \\
    Total $\Uplambda$CDM & --- & 0.29 & 1.14 & 11\% \\
    \hline\hline
\end{tabular}
\caption{A set of best-fit parameters for $\Lambda$CDM and DGP
cosmologies with $\Omega_\Lambda = 0$. `SN' represents fits to just
SN data; `CMB' are fits to the distance to the last-scattering
surface; `total' combine the two data sets. For $\Lambda$CDM, the
two fits are consistent, for pure DGP they are not, resulting in a
significantly worse overall fit. \label{t:bestfits}}
\end{centering}
\end{table}

\begin{figure}[htp]\begin{centering}
\includegraphics[width=10cm]{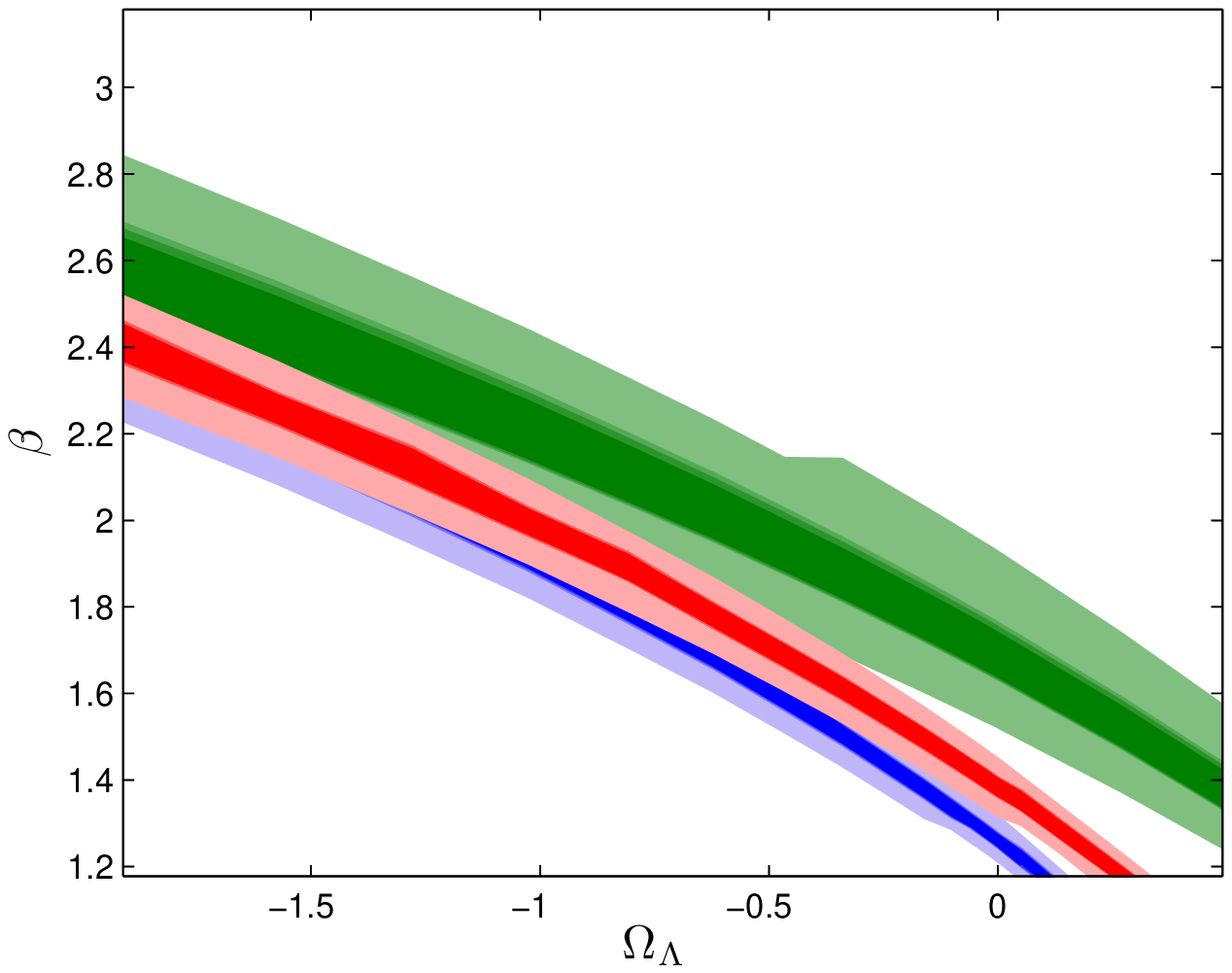} \\
\begin{tabular}{rrrr}
    \hline\hline
    \head{$\Omega_\Lambda$} & \head{$\beta\equiv r_\cs H_0$} &
        \head{$\Omega_\ms$} & \head{$\chi^2$}\\
    \hline
   $-10.0$ &    4.70 &    0.29 &  177 \\
   $-5.0$ &    3.47 &    0.29 &  177 \\
   $-2.0$ &    2.44 &    0.29 &  177 \\
   $-1.0$ &    1.98 &    0.29 &  179 \\
   $-0.5$ &    1.71 &    0.28 &  181 \\
       0&    1.38 &    0.28 &  188 \\
    0.2 &    1.23 &   0.27 &  197 \\
    0.5  & 0.99  &  0.29 & 228 \\
    \hline\hline
\end{tabular}
\caption{Plot showing the one sigma (dark color) and three sigma
(light color) range for best-fit values of $\beta$ for given values
of $\Omega_\Lambda$ for CMB distance (green), SN (blue) and combined
(red) For $\Omega_\Lambda$ close to 0 the preferred values for the
two data sets are significantly different, leading to a poor overall
fit. As $\Omega_\Lambda \rightarrow -\infty$ the preferred parameter
spaces increasingly overlap. In this regime DGP is indistinguishable
from GR. The table presents values of best-fit $\beta$'s for a
selection of $\Omega_\Lambda$ and the $\chi^2$'s of the respective
fits to combined CMB-distance and SN data. It can be clearly seen
that a positive bulk cosmological constant is strongly
excluded.\label{f:olbeta}}
\end{centering}\end{figure}

Using both the SN and WMAP data, with both $\Omega_\Lambda$ and
$\beta$ free, the maximum likelihood for the DGP cosmology is
attained for $\Omega_\Lambda = -7.3$ and $\beta = 4.1$. However,
since the range of admissible values of $\beta$ increases for more
negative $\Omega_\Lambda$, the likelihood for just $\Omega_\Lambda$
(marginalized over $\beta$) increases monotonically as
$\Omega_\Lambda$ attains lower values. Figure~\ref{f:olbeta} shows
the likelihood for $\beta$ given a particular value of
$\Omega_\Lambda$, while figure~\ref{f:omegaliktotal} shows the
likelihood for $\Omega_\Lambda$ marginalized over the supernova
absolute magnitude and $\beta$.  From the expansion history alone,
ordinary GR (corresponding to $\Omega_\Lambda \rightarrow -\infty$)
is preferred.

\subsection{Simulation Results}

\begin{figure}[htp]\begin{centering}
\includegraphics[width=10cm]{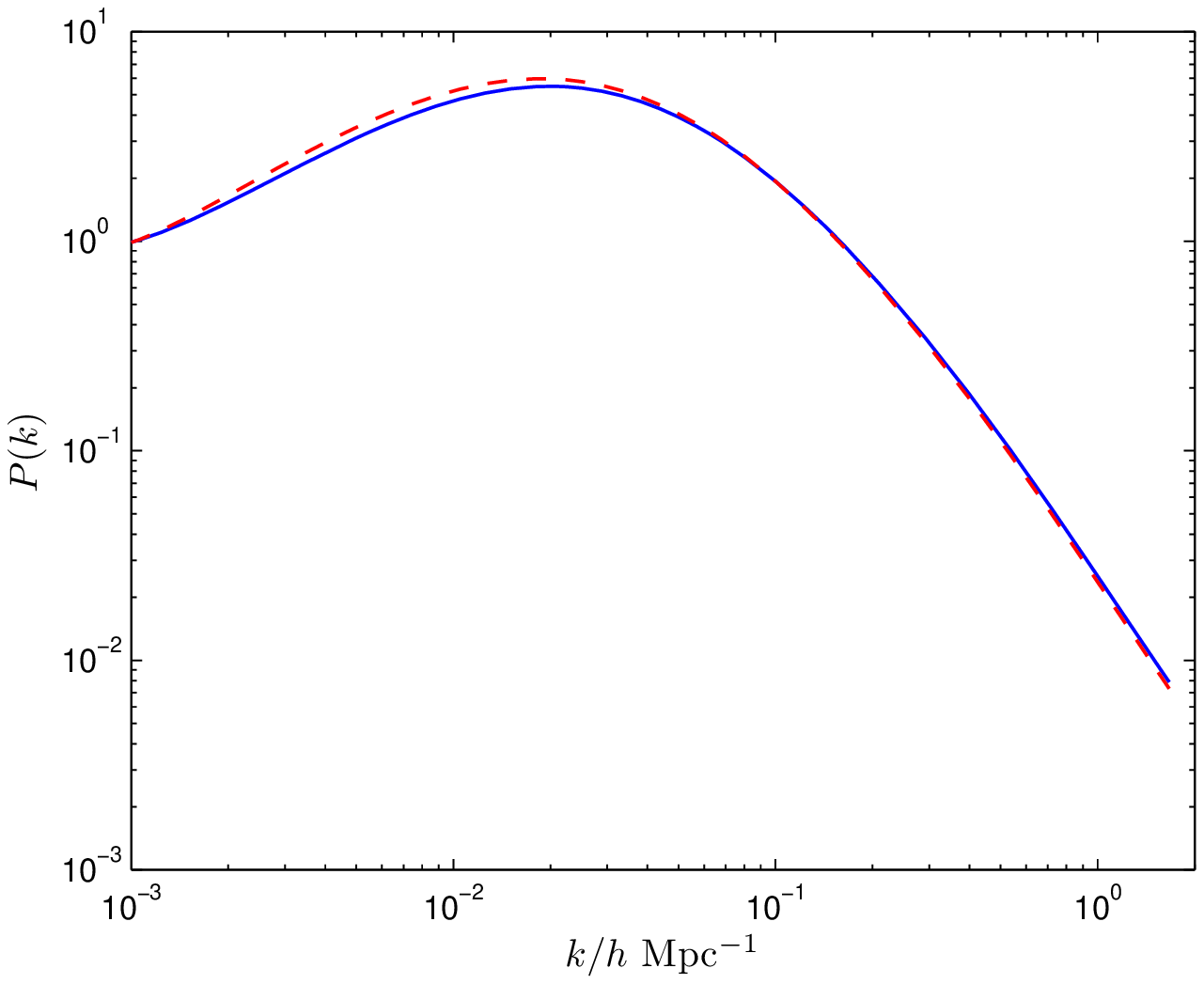}
\caption{Comparison of dark-matter power spectra for $\Uplambda$CDM
(solid line) and DGP ($\Omega_\Lambda = 0, \beta = 1.38$, dashed).
The spectra have been normalized to unity at $k = 10^{-3}
h$~Mpc$^{-1}$. There is excess power at large scales and a power
deficiency at low scales. \label{f:matpow}}
\end{centering}\end{figure}

\begin{figure}[htp]\begin{centering}
\includegraphics[width=10cm]{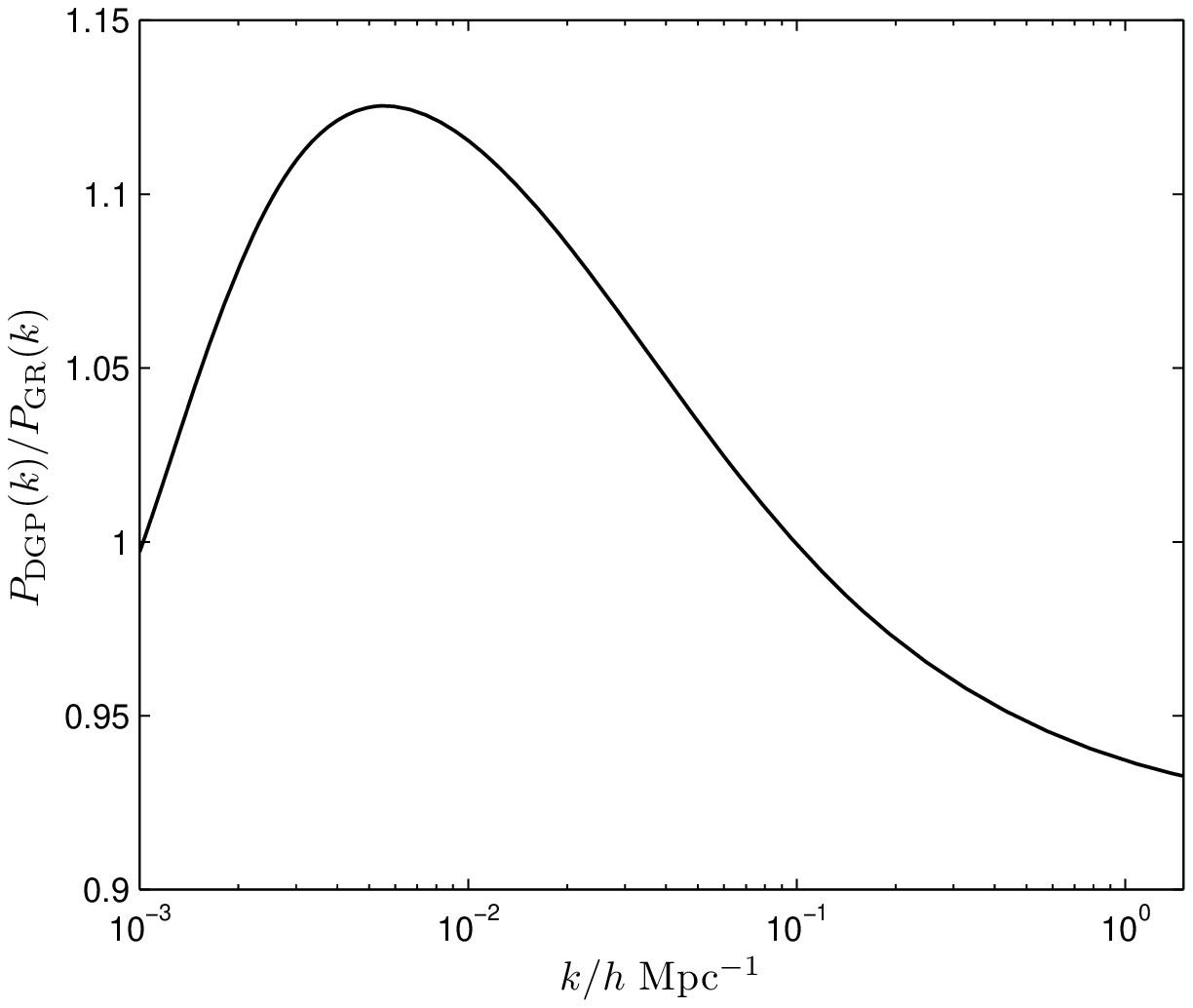}
\caption{Ratio of dark-matter power spectra for DGP ($\Omega_\Lambda
= 0, \beta = 1.38$) and $\Uplambda$CDM. Both spectra are initially
normalized to unity at $k=10^{-3} h$ Mpc$^{-1}$.
\label{f:matpowrat}}
\end{centering}\end{figure}

The matter power spectrum for DGP is slightly different than that
for $\Uplambda$CDM. We have found that there is excess power at
large scales and a deficiency of power at low scales. Results are
shown in Figures~\ref{f:matpow}~and~\ref{f:matpowrat}. This change
is a result of the different late-time evolution of the
gravitational potential and the change in the rate of growth of
density perturbations associated with it.

\begin{figure}[htp]\begin{centering}
\includegraphics[width=10cm]{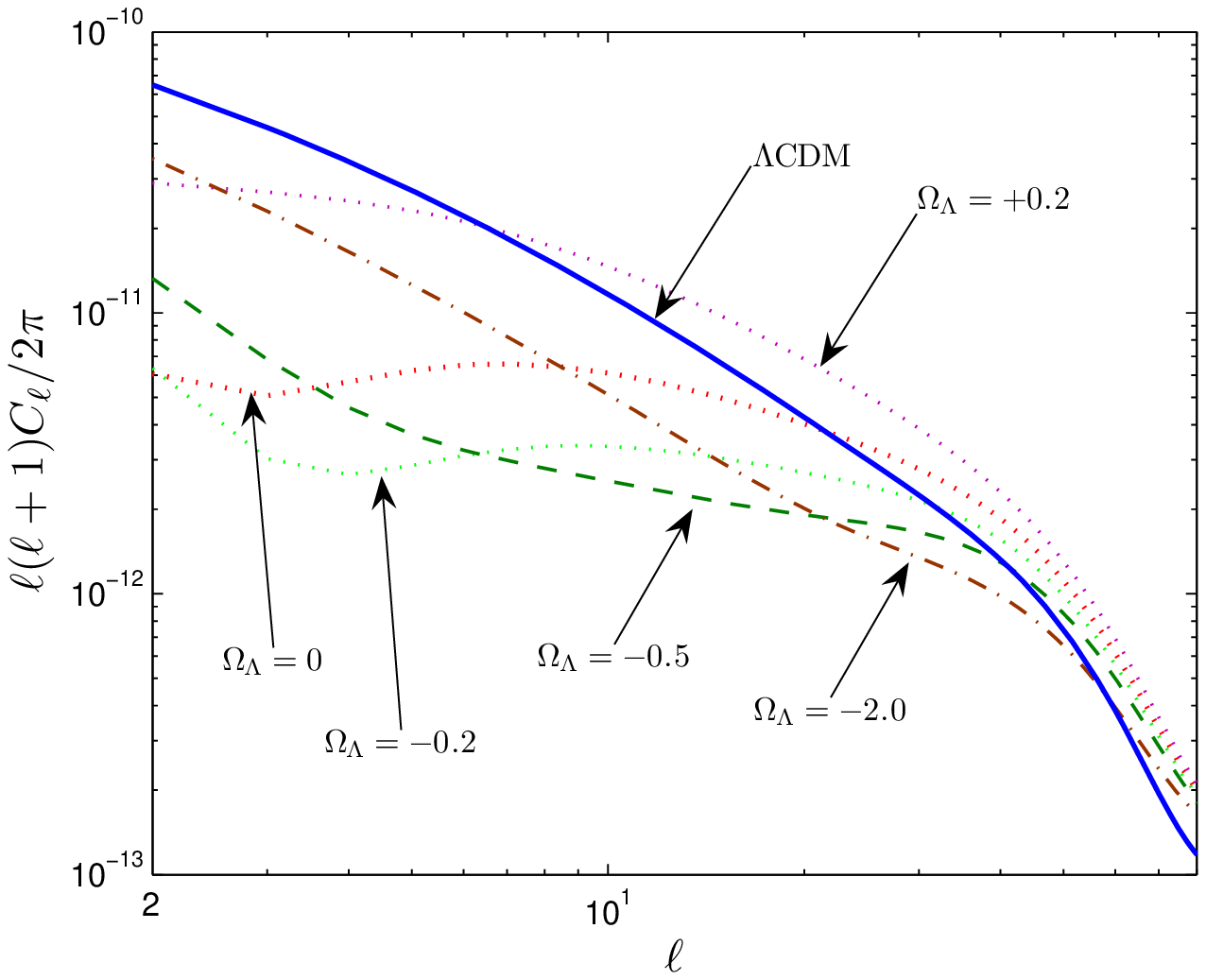}
\caption{Contribution to $C_\ell$ from ISW effect as calculated for
$\Uplambda$CDM and a range of DGP cosmologies. The effect is
significantly reduced at low multipoles in DGP. Making the bulk
cosmological constant more negative brings the effect back to
$\Uplambda$CDM levels. Making $\Omega_\Lambda$ more negative brings
the DGP power spectrum towards the $\Uplambda$CDM one.
\label{f:isw}}
\end{centering}\end{figure}

Using the sets of parameters presented in Figure~\ref{f:olbeta}, we
computed the ISW effect for the DGP model. We have found it to be
significantly reduced at low multipoles and to have a (very) slight excess
in the power for $\ell$ above 20 as compared to the concordance
$\Uplambda$CDM model. Making the bulk cosmological more negative
restores the behavior observed in $\Uplambda$CDM. A positive
$\Omega_\Lambda$ significantly increases the ISW effect. The results
are shown in Figure~\ref{f:isw}.

\subsection{Impact on CMB}

{}From the fits performed in \S\ref{s:fits}, we obtain a range of
values of $\beta$ for each $\Omega_\Lambda$ which satisfy the
constraints of the supernova data and the distance to the
last-scattering surface. In our subsequent analysis and simulations
we pick as $\beta$ the central values of the likelihood
distributions for each $\Omega_\Lambda$. A selection of
$\Omega_\Lambda$ and $\beta$ pairs is presented in the table
contained in Figure~\ref{f:olbeta}.

By requiring that the distance to the last scattering surface is
effectively fixed and by assuming the same
radiation-density-to-matter-density ratio as in the concordance
model, we ensure that the part of the CMB spectrum resulting from
plasma oscillations remains unchanged, despite the different theory
of gravity. We are thus able to take the $\Uplambda$CDM output of
CMBFAST and `replace' the ISW part of the power spectrum with the
new DGP calculation, provided we properly take into account the
cross-correlations between the ISW effect and the other
contributions to the CMB power spectrum.

Since the low multipoles are dominated by the SW and ISW effects, we
can compute both the ISW and SW effect in all cases and then assume
that the correlation between the two is equal to the correlation
between the ISW contribution and the rest of the signal. The
procedure is explained in detail in Appendix~\ref{s:correls}.

\begin{figure}[htp]\begin{centering}
\includegraphics[width=11cm]{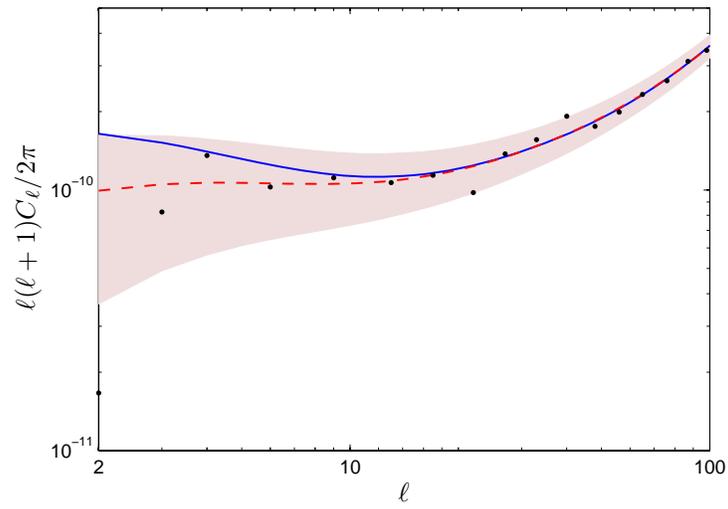}
\caption{Output of CMBFAST for concordance $\Uplambda$CDM (blue
solid line) and its modification for pure DGP ($\Omega_\Lambda =0$,
red dashed line). Shaded are represents cosmic variance for pure
DGP. WMAP measurements have been plotted in black. DGP reduces power
at the lowest multipoles, bringing the  power spectrum into slightly
better agreement with measurement. \label{f:cmb}}
\end{centering}\end{figure}

\begin{figure}[htp]\begin{centering}
\includegraphics[width=10cm]{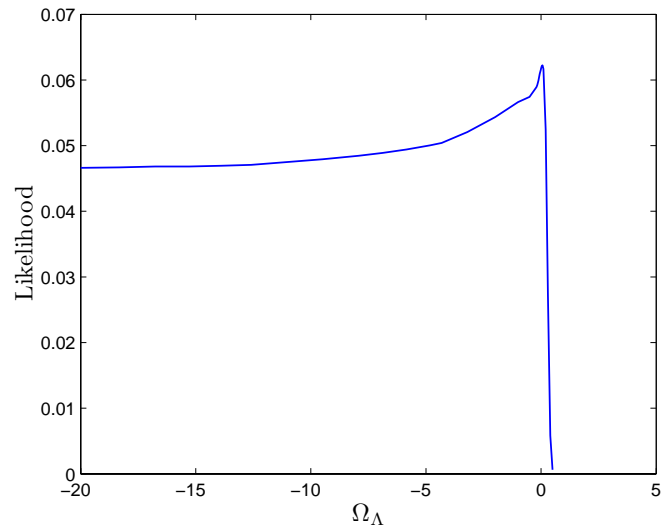}
\caption{Likelihood for a range of values of $\Omega_\Lambda$ as
implied solely by data from low-multipole WMAP measurements. The
preferred value is $\Omega_\Lambda = +0.06$ with the upper
half-maximum lying at lying at $+0.3$. Normalization is arbitrary.
\label{f:cmblik}}
\end{centering}\end{figure}

The results of performing the above procedure for models for DGP
cosmologies with a range of $\Omega_\Lambda$ (with $\beta$ as
implied by SNe Ia data and $d_\text{ls}$ as from WMAP) are presented
in Figure~\ref{f:cmb}. The total signal strength at low multipoles
is significantly reduced. The likelihood for $\Omega_\Lambda$ as
implied by the fit to the low-multipole WMAP data alone is presented
in Figure~\ref{f:cmblik}. For this data alone the maximum likelihood
is found at $\Omega_\Lambda = +0.06$, with a bulk cosmological
constant higher than that strongly excluded. Thus, the CMB data {\em
alone} slightly prefer pure DGP to $\Uplambda$CDM.

\begin{figure}[htp]\begin{centering}
\includegraphics[width=10cm]{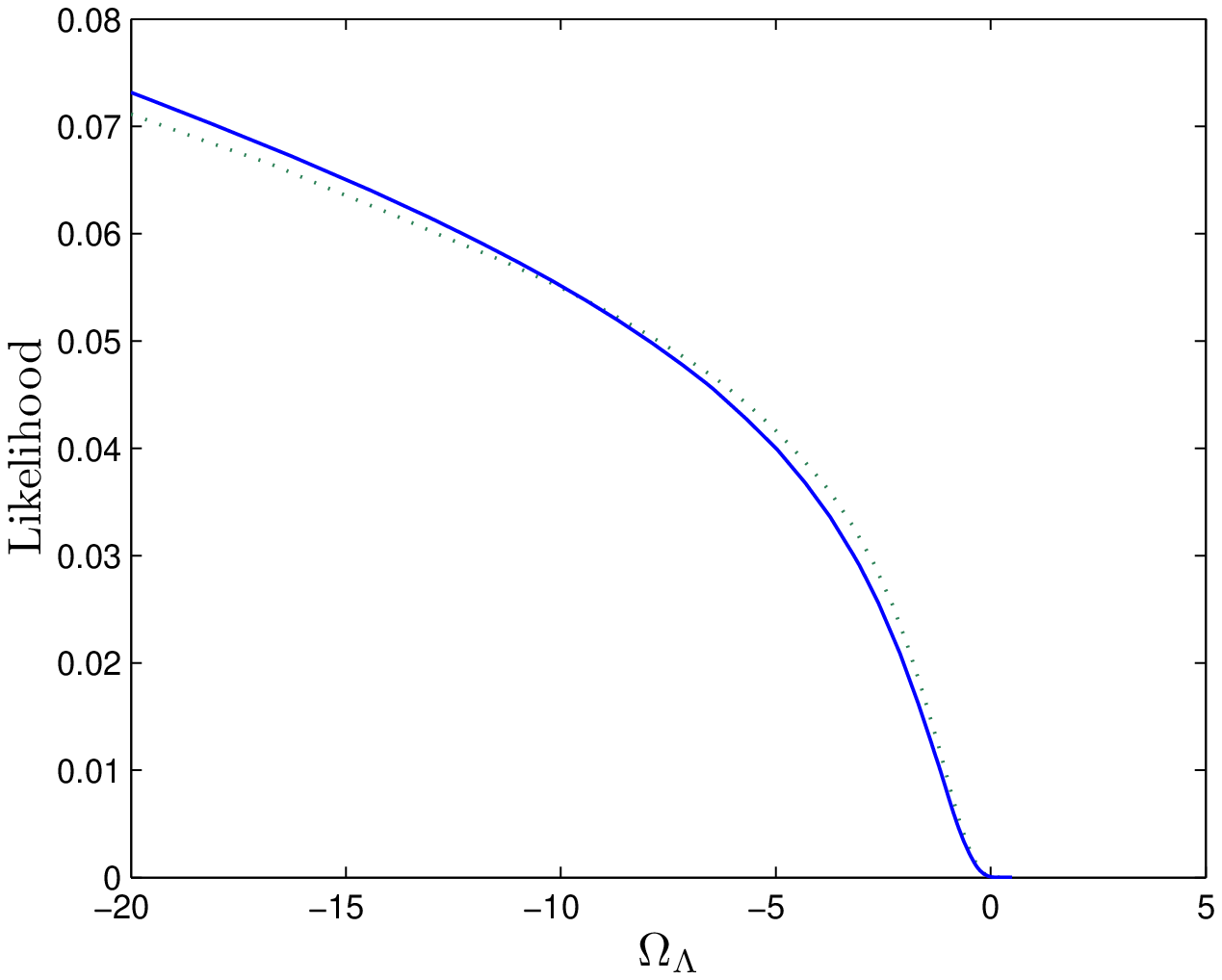}
\caption{Likelihood for a range of values of $\Omega_\Lambda$ as
implied by data: dotted line includes only CMB distance and SNe Ia.
Solid line is the modification to the likelihood caused by adding
constraints from low-multipole WMAP data. Despite the fact that WMAP
data prefers pure DGP, the preferred choice for the value of the
bulk cosmological constant is high enough to be excluded by the
other data. $\Uplambda$CDM is clearly preferred.
\label{f:omegaliktotal}}
\end{centering}\end{figure}

Finally, we combine the likelihoods obtained from the SNe Ia data,
and the CMB data, presented in Figure~\ref{f:cmblik} to obtain an
overall likelihood distribution for $\Omega_\Lambda$. This has been
presented in Figure~\ref{f:omegaliktotal}. Taking all experimental
data together does not change the conclusion that $\Lambda$CDM is
preferred to DGP: the better fit to low-$\ell$ multipole data of
WMAP is not enough to compensate for the inconsistency between fits
to CMB distance and supernova data.

\section{Conclusions}

We have performed a projection of the equations of the
Dvali--Gabadadze--Porrati modified theory of gravity in 5D bulk
(Minkowski, de Sitter and anti-de Sitter) onto a 4D brane embedded
in it. We have rederived a cosmological solution to the theory and
have derived equations governing the evolution of linear
perturbations. We find that the theory governing the linear
perturbations, in certain regimes, is one where $G$ is not a
constant, but is dependent on the average energy density in the
universe.

Using the new equations, we built a simple cosmological simulation
containing radiation and dark matter, driven by DGP gravity. We have
discovered that, in DGP cosmologies, the Newtonian potential
$\Phi$ exhibits a period of growth at late times, prior to its
decay; this is in contrast with GR, where the potential decays
once dark energy dominates. The precise details of this effect are a
function of the wavelength of the perturbation, and lead to an
altered transfer function and a changed dark-matter power spectrum,
with slightly higher power at large scales.

The impact of the change in the evolution of the potentials can be
seen through the Integrated Sachs-Wolfe effect. We simulate the CMB
anisotropy and demonstrate that in DGP cosmologies the ISW effect is
significantly weaker at low multipoles.

We constrain the parameters of the theory through a fitting
procedure using supernova data and the WMAP results and perform a
calculation of the likelihood function for the parameter space. We
find that assuming a flat cosmology, pure DGP with no bulk
cosmological constant is significantly worse when simultaneously fit
to supernova data and the distance to the last-scattering surface.
DGP fits better to low-$\ell$ multipole data from WMAP, owing to the
reduced ISW effect. However, this does not compensate for the poor
fit to the former, leading to the conclusion that GR and
$\Lambda$CDM are preferred by all the data available taken together.

\section{Acknowledgements}

We would like to thank Wayne Hu, Yong-Seon Song and Dragan Huterer
for fruitful discussions. This work was supported in part by the
U.S. Dept. of Energy contract DE-FG02-90ER-40560, the NSF grant
PHY-0114422, and the David and Lucile Packard Foundation. The KICP
is an NSF Physics Frontier Center.

\appendix
\section{Derivation of On-Brane Field Equations}
\label{s:deriv}

In this appendix we will replicate the work of
\cite{shiromizu:99proj,maeda:03dgpproj}, deriving the projected,
on-brane field equations for DGP gravity in 5D bulk.

We start off with Einstein's equation in a 5D manifold $\Mcal$
\begin{equation}
    G_{\mu\nu} = \kappa^2 T_{\mu\nu} \label{e:e} \,.
\end{equation}
The Greek indices run over all the dimensions, i.e.\ from 0 to 4.

We want to put in a 3-brane, $\Bcal$, (we will denote its position
by invoking a spacelike vector field $\chi$ in the neighborhood of
the brane, such that $\chi = 0$ will coincide with the position of
the brane. We want then to find the effective equation of motion for
gravity on $\Bcal$ itself.

Let us first take care of the LHS of \eqref{e:e}. We will first
define the induced metric on the 3-brane, $q_{\mu\nu}$:
\begin{equation}
    q_{\mu\nu} = g_{\mu\nu} - n_\mu n_\nu \,,
\end{equation}
where $g_{\mu\nu}$ is the metric on $\Mcal$, and $n^\mu$ is a
spacelike vector field in $\Mcal$, with unit norm, which is normal
to the brane at $\chi = 0$. Now we invoke the Gauss equation to
calculate the Riemann tensor on the 3-brane
\begin{equation}
    ^{(4)}R^\alpha_{\ph{\alpha}\beta\gamma\delta} =\
    ^{(5)}R^\mu_{\ph{\mu}\nu\rho\sigma} q^\alpha_{\ph{\alpha}\mu}
    q^\nu_{\ph{\nu}\beta} q^\rho_{\ph{\rho}\gamma}
    q^\sigma_{\ph{\sigma}\delta} + K^\alpha_{\ph{\alpha}\gamma}K_{\beta\delta}
    - K^\alpha_{\ph{\alpha}\delta}K_{\beta\gamma} \label{e:gauss}
\,,
\end{equation}
where $K_{\mu\nu} := q^\alpha_{\ph{\alpha}\mu}
q^\beta_{\ph{\beta}\nu} \nabla_\alpha n_\beta$ is the extrinsic
curvature.

In order to simplify our notation, we are going to define the
perpendicular subscript to signify
$A_{\mu\perp}:=A_{\mu\nu}n^{\nu}$.

Contracting $\alpha$ and $\gamma$ in \eqref{e:gauss} we obtain the
4D Ricci tensor:
\begin{equation}
    ^{(4)}R_{\beta\delta} =\ ^{(5)}R_{\nu\sigma} q^\nu_{\ph{\nu}\beta}
    q^{\sigma}_{\ph{\sigma}{\delta}} -
    \ ^{(5)}R^\perp_{\ph{\perp}\nu\perp\sigma} q^\nu_{\ph{\nu}\beta}
    q^\sigma_{\ph{\sigma}\delta} + KK_{\beta\delta} -
    K_\beta^{\ph{\beta}\alpha}K_{\delta\alpha} \,.
\end{equation}
and, then, the 4D Ricci scalar:
\begin{align}
    ^{(4)}R :=\ ^{(4)}R_{\mu\nu}g^{\mu\nu} &= \
    ^{(5)}R_{\mu\nu}q^{\mu\nu} -\ ^{(5)}R^{\perp}_{\ph{\perp}\mu\perp\nu}
    q^{\mu\nu} + K^2 -K_\mu^{\ph{\mu}\nu} K^\mu_{\ph{\mu}\nu} \\
&=\ ^{(5)}R -\ ^{(5)}R_{\perp\perp} -\
    ^{(5)}R^{\perp\ph{\beta\perp}\beta}_{\ph{\perp}\beta\perp} +\
    \underbrace{^{(5)}R^\perp_{\ph{\perp}\perp\perp\perp}}_{=0} + K^2 -
K_{\alpha\beta}  K^{\alpha\beta} \,. \notag
\end{align}

An observer on the brane will only be able to measure the 4D Ricci
tensor and scalar, and not their five-dimensional counterparts, so
will naturally \emph{define} the 4D Einstein tensor as:
\begin{align}
^{(4)}G_{\beta\delta} &:=\ ^{(4)}R_{\beta\delta} -
\frac{1}{2}q_{\beta\delta} \ ^{(4)}R = \label{e:g4} \\
&=\ ^{(5)}R_{\mu\nu} q^\mu_{\ph{\mu}\beta} q^\nu_{\ph{\nu}\delta} -\
^{(5)}R^\perp_{\ph{\perp}\mu\perp\nu} q^\mu_{\ph{\mu}\beta}
q^\nu_{\ph{\nu}\delta} - \frac{1}{2} g_{\mu\nu}
q^\mu_{\ph{\mu}\beta} q^\nu_{\ph{\nu}\delta} \left(^{(5)}R -\
^{(5)}R_{\perp\perp} -\
^{(5)}R^{\perp\ph{\beta\perp}\alpha}_{\ph{\perp}\alpha\perp}
\right) \notag \\
&\qquad + KK_{\beta\delta} - K_\beta^{\ph{\beta}\alpha}
K_{\delta\alpha} - \frac{1}{2} q_{\beta\delta} \left( K^2 -
K_{\alpha\beta} K^{\alpha\beta} \right) \,. \notag
\end{align}

We can simplify this by:
\begin{align}
    ^{(5)}R_{\perp\perp} = g^{\alpha\beta}n^\gamma\
     ^{(5)} R_{\beta\gamma\alpha\perp} =
    g^{\alpha\beta}n_\gamma\, ^{(5)}R^\gamma_{\ph{\gamma}\beta\perp\alpha}
    =\ ^{(5)}R^{\perp\ph{\beta\perp}\beta}_{\ph{\perp}\beta\perp}
\,.
\end{align}
and also expanding the Riemann tensor in terms of the Riccis and the
Weyl tensor in 5D \cite[(3.2.28)]{wald:grbook}:
\begin{align}
    &q^\alpha_{\ph{\alpha}\mu} q^\beta_{\ph{\alpha}\nu} n^\rho n^\sigma\ ^{(5)}R_{\rho\alpha\sigma\beta}
    = q^\alpha_{\ph{\alpha}\mu} q^\beta_{\ph{\alpha}\nu} n^\rho
    n^\sigma \left( \frac{1}{3} \left(g_{\rho\sigma}\ ^{(5)}R_{\beta\alpha} -
    g_{\rho\beta}\ ^{(5)}R_{\sigma\alpha} - g_{\alpha\sigma}\ ^{(5)}R_{\beta\rho} + \right. \right. \\
& \left. \left. \qquad + g_{\alpha\beta}
    \ ^{(5)}R_{\sigma\rho}\right) -  \frac{1}{12}  \left( g_{\rho\sigma}
    g_{\beta\alpha} - g_{\rho\beta} g_{\sigma\alpha} \right)\ ^{(5)}R +
    \ ^{(5)}C_{\rho\alpha\sigma\beta} \right) = \notag \\
&= \frac{1}{3} \left(\ ^{(5)}R_{\alpha\beta}
q^\alpha_{\ph{\alpha}\mu}
    q^\beta_{\ph{\alpha}\nu} + q_{\mu\nu}\ ^{(5)}R_{\perp\perp} \right) -
    \frac{1}{12} q_{\mu\nu} \ ^{(5)}R + \underbrace{\ ^{(5)}C_{\perp\alpha\perp\beta} q^\alpha_{\ph{\alpha}\mu}
    q^\beta_{\ph{\alpha}\nu}}_{=: E_{\mu\nu}} \, . \notag
\end{align}
We can also manipulate the 5D Einstein equation \eqref{e:e} to
obtain:
\begin{align}
    ^{(5)}R &= -\frac{2}{3} \kappa^2 T \\
    ^{(5)}R_{\rho\sigma} &= \kappa^2 \left( T_{\rho\sigma}
    -\frac{1}{3} g_{\rho\sigma} T\right) \\
    R_{\perp\perp} &= \kappa^2 \left( T_{\perp\perp} - \frac{1}{3}
        T \right) \,.
\end{align}
Using all of the above in \eqref{e:g4}, we obtain:
\begin{align}
    ^{(4)}G_{\mu\nu} &=\ ^{(5)}G_{\rho\sigma} q_\mu^{\ph{\mu}\rho}
    q_\nu^{\ph{\nu}\sigma} +\, ^{(5)}R_{\perp\perp}
    q_{\mu\nu} + KK_{\mu\nu} - K_\mu^{\ph{\mu}\rho} K_{\nu\rho} -
    \frac{1}{2}q_{\mu\nu} \left(K^2 - K^{\alpha\beta}K_{\alpha\beta} \right) -
    \label{e:brane} \notag\\
&\qquad - \frac{1}{3}\, ^{(5)}R_{\alpha\beta}
    q^\alpha_{\ph{\alpha}\mu} q^\beta_{\ph{\beta}\nu} - \frac{1}{3}
    q_{\mu\nu}\, ^{(5)}R_{\perp\perp} + \frac{1}{12} q_{\mu\nu}\,^{(5)}R
    - E_{\mu\nu} =\notag\\
&= \frac{2\kappa^2}{3} \left( T_{\rho\sigma} q^\rho_{\ph{\rho}\mu}
    q^\sigma_{\ph{\sigma}\nu} + q_{\mu\nu}
    \left(T_{\perp\perp}-\frac{1}{4}T\right) \right) +KK_{\mu\nu} -
    K_\mu^{\ph{\mu}\rho}K_{\nu\rho} - \notag \\
&\qquad -\frac{1}{2} q_{\mu\nu}\left(K^2
-K^{\alpha\beta}K_{\alpha\beta} \right) - E_{\mu\nu} \,.
\end{align}

Now let's take care of the energy-momentum tensor. It has
contributions from the bulk (which we will limit to just a bulk
cosmological constant, $\Lambda$) as well as the brane contributions.
One of the contributions in the brane is the radiatively induced DGP
4-gravity term; the action here is just the usual $\mu^{-2}
\delta(\chi)\int \! \ds^4x \, ^{(4)}R$, restricted to the
brane, the variation of which gives the 4D Einstein tensor. The
stress-energy tensor has the form:
\begin{equation}
    T_{\mu\nu} = -\Lambda g_{\mu\nu} + \underbrace{\left( -\lambda q_{\mu\nu} +
    \tau_{\mu\nu} - \mu^{-2} \,^{(4)}G_{\mu\nu} \right)}_{=:S_{\mu\nu}}
\delta(\chi) \,.
\end{equation}

We are going to perform our calculation just off the brane. The
energy-momentum tensor there consists of just the bulk cosmological
constant. The extrinsic curvature can be determined from the Israel
junction conditions, one of the formulations of which allows us to
compute the jump in the metric and extrinsic curvature across a thin
shell of non-zero momentum-energy \cite{israel:66junc}:
\begin{align}
    q_{\mu\nu}^+-q_{\mu\nu}^- =: [q_{\mu\nu}] &= 0 \\
    [K_{\mu\nu}] &= -\kappa^2
    \left(S_{\mu\nu}-\frac{1}{3}q_{\mu\nu}S \right)
\end{align}
Assuming that the universe is symmetric about the brane allows us to
obtain the values of the extrinsic curvature explicitly,
$K_{\mu\nu}^+=-K_{\mu\nu}^- = \frac{1}{2}[K_{\mu\nu}]$ and $K^+=
\frac{1}{6}\kappa^2 S$. Substituting into \eqref{e:brane} and making
4D subscripts implicit:
\begin{align}
    G_{\mu\nu} &= -\frac{\kappa^2}{2} \Lambda q_{\mu\nu} -
    \frac{\kappa^4}{12} S\left(S-\frac{1}{3} q_{\mu\nu} S\right) -
    \frac{\kappa^4}{4}
    \left(S\ix{_\mu^\sigma}-\frac{1}{3}q\ix{_\mu^\sigma}S\right)
    \left(S_{\nu\sigma}-\frac{1}{3}q_{\nu\sigma}S\right)- \notag \\
&\qquad -\frac{\kappa^4}{8} q_{\mu\nu}\left(\frac{1}{9}S^2 -
    \left(S_{\mu\nu}-\frac{1}{3}q_{\mu\nu}S\right)^2\right) \,.
\end{align}
We then substitute for $S_{\mu\nu}$ from its definition and, after a
bit of algebra, obtain:
\begin{equation}
   \left(1+\frac{\lambda\kappa^4}{6\mu^2}\right)G_{\mu\nu} =
   -\left(\frac{\kappa^2}{2}\Lambda +\frac{\kappa^4\lambda^2}{12}\right) q_{\mu\nu}+ \frac{\lambda
   \kappa^4}{6}\tau_{\mu\nu} + \kappa^4\pi_{\mu\nu} +
   \frac{\kappa^4}{\mu^4} \gamma_{\mu\nu} + \frac{\kappa^4}{\mu^2}
   \xi_{\mu\nu} - E_{\mu\nu} \label{e:me} \,.
\end{equation}
where the new tensors are quadratic in $G_{\mu\nu}$ and
$\tau_{\mu\nu}$ and are defined as:
\begin{align}
    \pi_{\mu\nu} &= \frac{1}{12}\tau \tau_{\mu\nu} -
    \frac{1}{4}\tau\ix{_\mu^\alpha}\tau_{\nu\alpha} +
    \frac{1}{8}q_{\mu\nu} \left(\tau_{\alpha\beta}\tau^{\alpha\beta}-
    \frac{1}{3}\tau^2 \right) \label{e:pi} \\
\gamma_{\mu\nu} &= \frac{1}{12}G G_{\mu\nu} -
    \frac{1}{4}G\ix{_\mu^\alpha}G_{\nu\alpha} +
    \frac{1}{8}q_{\mu\nu}\left( G_{\alpha\beta}G^{\alpha\beta}-
    \frac{1}{3}G^2\right) \label{e:gamma} \\
\xi_{\mu\nu} &= -\frac{1}{12}\left(\tau G_{\mu\nu}+\tau_{\mu\nu}
    G\right) +\frac{1}{4} \left(\tau\ix{_\mu^\alpha} G_{\alpha\nu} +
        G\ix{_\mu^\alpha}\tau_{\nu\alpha}\right) - \frac{1}{4}q_{\mu\nu}
    \left( \tau_{\alpha\beta} G^{\alpha\beta} - \frac{1}{3}\tau G
    \right) \label{e:xi} \\
E_{\mu\nu} &=
C\ix{^\perp_\alpha_\perp_\beta}q\ix{^\alpha_\mu}q\ix{^\beta_\nu} \,.
\end{align}
These quadratic tensors are related to $f_{\mu\nu}$ of (\ref{fmunu})
by
\begin{equation}
  f_{\mu\nu} = \mu^4 \pi_{\mu\nu} + \mu^2 \xi_{\mu\nu} + \gamma_{\mu\nu}\, .
\end{equation}

\section{Modifying CMBFAST data}
\label{s:correls}

This appendix explains how CMBFAST data was modified to reflect the
expected DGP signal.

Our cosmological simulation was used to calculate the contribution
to the CMB power spectrum from the ISW and SW effects, as well as
the total of the two (taking into account the cross-correlation). We
then defined the ISW-SW cross-correlation factor as
\begin{equation}
    R := \frac{C_\ell^\text{ISW-SW}}{\sqrt{C_\ell^\text{ISW-ISW}C_\ell^\text{SW-SW}}} =
\frac{C_\ell^\text{total} - C_\ell^\text{SW-SW} -
C_\ell^\text{ISW-ISW}}{2\sqrt{C_\ell^\text{ISW-ISW}C_\ell^\text{SW-SW}}}
\end{equation}

The $R$ thus defined was then assumed to be identical to the
cross-correlation factor between the ISW effect and the total
CMBFAST signal. This is a very good approximation at the lowest
multipoles. With this assumption we can compute the non-ISW part of
the power spectrum by solving for $C_\ell^{\sim\text{ISW}}$ in

\begin{equation}
    C_\ell^\text{CMBFAST} \approx C_\ell^{\sim\text{ISW}} +
2\sqrt{C_\ell^{\sim{\text{ISW}}}C_\ell^\text{ISW-ISW}} R +
C_\ell^\text{ISW-ISW} \label{e:corr}
\end{equation}
Finally, to compute the power spectrum for DGP cosmologies, we take
the values of $R$ and $C_\ell^\text{ISW-ISW}$ calculated for the
particular model and use them in equation \eqref{e:corr}. This
output is presented in Figure~\ref{f:cmb}.

\bibliographystyle{JHEP-2}
\bibliography{DGPPaper_bib}
\end{document}